\documentclass[aps,amsfonts,prl,nofootinbib,tightenlines,showpacs,superscriptaddress,twocolumn]{revtex4}

\usepackage{bm}
\usepackage{graphicx}
\usepackage{ulem,color}
\usepackage{amsmath}

\newcommand{\sech}{{\rm sech}}

\begin{document}

\title{Attractive Electromagnetic Casimir Stress on a Spherical
Dielectric Shell}

\author{N. Graham}
\affiliation{Department of Physics, Middlebury College
Middlebury, VT 05753, USA}

\author{M. Quandt}
\affiliation{Institute for Theoretical Physics, University of T\"ubingen,
D--72076 T\"ubingen, Germany}

\author{H. Weigel}
\affiliation{Physics Department, Stellenbosch University,
Matieland 7602, South Africa}

\begin{abstract}
Based on calculations involving an idealized boundary condition, it has 
long been assumed that the stress on a spherical conducting shell is 
repulsive. We use the more realistic case of a Drude dielectric to show 
that the stress is attractive, matching the generic behavior of Casimir 
forces in electromagnetism. We trace the discrepancy between these 
two cases to interactions between the electromagnetic quantum
fluctuations and the dielectric material.
\end{abstract}

\maketitle

After discovering the quantum-mechanical force between uncharged
conductors that bears his name, Casimir proposed using this force to
model the electron as a uniformly charged spherical shell whose size
is fixed by balancing its attractive Casimir stress against its
electrostatic repulsion \cite{CasimirShell}.  Subsequent calculations
\cite{Boyer:1968uf,Milton:1978sf}, however, found a repulsive rather
than attractive stress in this geometry, and in any case quantum
electrodynamics (QED) provides a fuller description of the electron.  
Since then, this result has been extrapolated outside the realm of fundamental
physics, with the result that the existence of a repulsive Casimir
stress on a conducting sphere has been taken as a standard result in
the field, even though it is very much at odds with the attractive
\emph{force} between two hemispheres \cite{Kenneth06} and is not
robust against the infinitesimal deformation or nonzero thickness of
the spherical shell.

In this Letter we argue that, when applied to a mesoscopic conducting
shell, Casimir's original picture of an attractive force was indeed correct.
At the center of the difficulty in computing Casimir stresses are the
divergences --- more precisely, the dependences on the short-distance
cutoff --- inherent in these calculations.  While a fundamental theory
of the electron can postulate any model for these short-distance
effects, a calculation relevant to mesoscopic materials does not have
this freedom.  Instead, we employ the standard Drude model for metals
(though the simpler plasma model yields similar results). The field 
theory cutoff must be imposed at distances shorter than any other 
scale in the problem; in particular, the cutoff must be at scales 
shorter than the plasma wavelength, at which the material no longer 
acts as a perfect conductor.  As a result, it is essential that these 
two limits are taken in the correct order, which implies that fluctuations 
at the scale of the cutoff should always see the material as transparent.

While we will use a generic dielectric to model the shell material, our
results agree qualitatively with results obtained from more specific
models, such as a carbon nanostructure \cite{Barton} or a
``fish-eye'' medium \cite{PhysRevD.84.081701}.  Our findings are also in
agreement with the work of Deutsch and Candelas
\cite{Deutsch:1978sc,Candelas1982241}, who showed that
divergences in Casimir stresses arise from surface
counterterms \cite{Symanzik:1981wd} that cannot be removed by
renormalization, and with explicit calculations in scalar models
\cite{Graham:2002fw,Graham:2002xq,Graham:2003ib}.  Finally, it should
also be noted that the Casimir energy of an idealized boundary can be
of interest for the mathematical ``Weyl problem'' of the relationship
between eigenvalue spectra and geometry
\cite{Abalo:2012jz,Kolomeisky:2010eg}.
This approach draws on classic results relating the shape of a boundary to
the density of scattering states \cite{Balian:1970fw,Balian:1976za},
but does not make direct contact with a physically measurable stress.

\medskip
We model the shell as a space- and frequency-dependent Drude dielectric,
\begin{equation}
\epsilon_k(r) = 1+\frac{(2\pi)^2}{-(\lambda_p k)^2 +
\frac{\pi}{\sigma_p} \sqrt{-k^2}} \, p(r)
\label{101}
\end{equation}
where $p(r)$ is a spherically symmetric profile function that goes to
zero for $r\to\infty$, $\lambda_p$ is the plasma wavelength,
$\sigma_p$ is the conductivity, and there is no free charge.  
In the present study we mainly aim at a proof of principle
calculation rather than investigating a specific material. In this
framework the conductivity term merely plays the role of an infra-red
regulator that is typically required in QED calculations.

One question that comes up immediately is why, in the limit
where the dielectric approaches a perfect conductor,
our result should differ from the repulsive stress that Boyer obtained
by considering ideal boundary conditions.  Indeed,
Ref.~\cite{Bordag:2008rc} has shown how the Boyer result emerges as
the limit of a plasma model shell for large radii and/or plasma
frequencies (note that that paper also finds an attractive stress, in
agreement with our results, in the opposite limit).  We will see that
the attractive stress we find is due to an additional contribution
arising from the material properties of the shell, which overcomes the
standard repulsion at large radii and plasma frequencies.

In Eq.~(\ref{101}), we have given the dielectric function for 
real frequencies, where it has an imaginary part. Of course, 
vacuum fluctuations should not lead to dissipation. The imaginary 
part of the dielectric constant is really an effective model of
the interaction of the electro--magnetic field with the atoms of
the material. In appendix A of Ref.~\cite{Rahi:2009hm} it has 
been shown that by starting from a full particle model of the
material, in which case the energy is manifestly real, one can 
derive the vacuum polarization energy as the analytic continuation 
of the dielectric function to imaginary frequencies using a path 
integral approach.  As a result, this calculation yields a 
purely real vacuum polarization energy. It is exactly this analytic 
continuation that we perform for the calculation of the vacuum 
polarization energy (see Eq.~(\ref{eqn:renormalized}) below).

We decompose the quantum fluctuations by frequency $\omega = ck$ and work
in units where $\hbar = c = 1$.  Because we maintain spherical symmetry 
and parity, the transverse electric (TE) and transverse magnetic (TM) 
modes decouple.  As a result, we can use the method of Ref.~\cite{Johnson99} 
to reduce each channel to a scalar scattering problem. For the TE mode, 
we parameterize the electric field as 
\begin{equation}
\bm{E}_k(\bm{r}) = k \nabla \times [\varphi_k(\bm{r}) \bm{r}] = -k\bm{r} \times
\nabla \varphi_k(\bm{r}) \,,
\label{eqn:goodE}
\end{equation}
which obeys
\begin{equation}
\nabla \cdot \left(\epsilon_k(r) \bm{E}_k(\bm{r})\right) = \nabla
\cdot \left[k \nabla \times \left(\epsilon_k(r) \varphi_k(\bm{r})
\bm{r}\right)\right] = 0
\end{equation}
and solves the Maxwell equation
\begin{equation}
\nabla \times \nabla \times \bm{E}_k(\bm{r}) = k^2 \epsilon_k(r)
\bm{E}_k(\bm{r})
\label{eq:Maxwell}
\end{equation}
if $-\nabla^2 \varphi_k(\bm{r}) = k^2 \epsilon_k(r) \varphi_k(\bm{r})$.  
We solve the latter equation using separation of variables
\begin{equation}
\varphi_{k,\ell m}(\bm{r}) = \frac{1}{\sqrt{\ell(\ell+1)}}
Y_{\ell}^m(\theta,\phi) \frac{1}{r}f_{k,\ell}(r) \,.
\end{equation}
with $\ell=1,2,3\ldots$ and
\begin{equation}
-f_{k,\ell}''(r) + \frac{\ell(\ell+1)}{r^2}f_{k,\ell}(r) - k^2 \epsilon_k(r)
f_{k,\ell}(r) = 0 \,.
\label{eq:x8}
\end{equation}
Here, the prime denotes a derivative with respect to $r$.

For the TM mode, we take 
\begin{equation}
\bm{B}_k(\bm{r}) = \frac{k}{i} \nabla \times [\phi_k(\bm{r}) \bm{r}] \,,
\label{eqn:badE}
\end{equation}
which obeys $\nabla \cdot \bm{B}_k(\bm{r})=0$.  Then we have
\begin{equation}
\label{eqn:tm}
\bm{E}_k(\bm{r}) = \frac{i}{k\epsilon_k(r)} \nabla \times \bm{B}_k(\bm{r})
= \frac{1}{\epsilon_k(r)} \nabla \times
\nabla \times [\phi_k(\bm{r}) \bm{r}] \,,
\end{equation}
which obeys $\nabla \cdot [\epsilon_k(r) \bm{E}_k(\bm{r})]=0$. Again 
we use separation of variables to write the solution in the form
\begin{equation}
\phi_{k,\ell m}(\bm{r}) = 
\frac{\sqrt{\epsilon_k(r)}}{\sqrt{\ell(\ell+1)}} Y_{\ell}^m(\theta,\phi) 
\frac{1}{r} g_{k,\ell}(r) \,.
\label{eq:x9}
\end{equation}
The Maxwell equation 
$\nabla\times\bm{B}_k(\bm{r})=-ik\epsilon_k(\bm{r})\bm{E}(\bm{r})$
then gives
\begin{eqnarray}
k^2 g_{k,\ell}(r) &=& -g_{k,\ell}''(r) +\frac{\ell(\ell+1)}{r^2}
g_{k,\ell}(r) +
\\ &&
\left[k^2(1-\epsilon_k(r)) + \frac{3 \epsilon_k'(r)^2}{4 \epsilon_k(r)^2}
- \frac{\epsilon_k''(r)}{2 \epsilon_k(r)} \right] g_{k,\ell}(r)
\,, \nonumber
\end{eqnarray}
where in Eq.~(\ref{eq:x9}) we have introduced the scaling factor $\sqrt{\epsilon_k(r)}$ 
to ensure that the single--particle wave equation is Hermitian.

A single frequency mode contributes the energy density
$$
\frac{1}{2}\epsilon_k(r)\bm{E}_k(\bm{r})^2
+\frac{1}{2}\bm{B}_k(\bm{r})^2\,.
$$
Integration over space
and use of the wave equation yields an expression 
for the total energy that takes the same form in both channels,
\begin{eqnarray}
E &=& \int_0^\infty dk \sum_{\ell=1}^\infty (2\ell+1) 
\frac{k}{\pi} \int_0^\infty dr  \left(\epsilon_k(r)
|\psi_{k,\ell}(r)|^2 
\phantom{\psi^{(0)}} \right. \cr && \hspace{1cm} - \left.
|\psi_{k,\ell}^{(0)}(r)|^2 \right)\,,
\label{eqn:energy}
\end{eqnarray}
where $\psi$ is the normalized physical scattering solution for either 
$f$ or $g$ and we have summed over all fluctuating modes. In this 
expression, we have subtracted the energy in the absence of the shell 
using the corresponding free wavefunction $\psi_{k,\ell}^{(0)}(r)$.
Eq.~(\ref{eqn:energy}) is the formal field theory result
for the unrenormalized energy of the photon quantum fluctuations. 
To prepare it for renormalization and to put it in a form better 
suited for numerical evaluation we first express it as
\begin{eqnarray}
E &=& \int_0^\infty dk \sum_{\ell=1}^\infty (2\ell+1) 
\frac{k}{\pi} \int_0^\infty dr \label{eqn:energy2}
\\ &&  \hspace{-0.5cm}\times
\left( |\psi_{k,\ell}(r)|^2-|\psi_{k,\ell}^{(0)}(r)|^2 
+\left(\epsilon_k(r)-1\right) |\psi_{k,\ell}(r)|^2 
\right)\,.
\nonumber
\end{eqnarray}
In standard computations of Casimir (or vacuum polarization)
energies due to interactions with static backgrounds, the last term 
is absent. Fortunately this modification is numerically well-behaved 
because the radial integral is over a finite domain. The radial integral
$\int dr \left( |\psi_{k,\ell}(r)|^2-|\psi_{k,\ell}^{(0)}(r)|^2 \right)$
is proportional to the change of the density of states induced by the 
interaction with the background. This integral is much more difficult to 
compute numerically because of delicate cancellations between oscillating 
functions at large radii. In ordinary potential scattering theory this 
obstacle is solved by relating the change in the density of states to
the momentum derivative of the phase shift $\delta_\ell(k)$. We still can 
use that approach to compute the troublesome radial integral, but have to 
account for changes due to the energy dependences of the potentials in 
Eqs.~(\ref{eq:x8}) and~(\ref{eq:x9}). In particular, the Jost function analysis of
Ref.~\cite{Graham:2002xq} now yields
\begin{eqnarray}
\frac{1}{\pi} \frac{d\delta_\ell}{dk} 
&=& \frac{2}{\pi} \int_0^\infty dr\left[ 
\frac{1}{2k} \frac{d}{dk} \left(k^2 - V_k(r)\right)
|\psi_{k,\ell}(r)|^2 
\right. \cr && \hspace{1.5cm} \left. \phantom{\frac{1}{1}}
- |\psi_{k,\ell}^{(0)}(r)|^2 \right]  \cr
&=&\frac{2}{\pi} \int_0^\infty dr\left[ 
|\psi_{k,\ell}(r)|^2- |\psi_{k,\ell}^{(0)}(r)|^2 
\right. \cr && \hspace{1.0cm} \left. \phantom{\frac{1}{1}}
-\frac{1}{2k} |\psi_{k,\ell}(r)|^2 \frac{d}{dk} V_k(r)
\right]
\label{eqn:goodwronk}
\end{eqnarray}
The potential, $V_k(r)$, is $\displaystyle V^{\hbox{\tiny TE}}_k(r) = 
k^2 (1-\epsilon_k(r))$ and $\displaystyle V^{\hbox{\tiny TM}}_k(r) = k^2
(1-\epsilon_k(r)) + \frac{3 \epsilon_k'(r)^2}{4 \epsilon_k(r)^2} -
\frac{\epsilon_k''(r)}{2 \epsilon_k(r)}$ for the 
TE and TM channels, respectively.
By contrast, for ordinary potential scattering, the factor of
$\epsilon_k(r)$ would be absent from Eq.~(\ref{eqn:energy}) and the
potential would be $k$-independent, meaning that the $r$ integrands in
Eqs.~(\ref{eqn:energy}) and (\ref{eqn:goodwronk}) would coincide.

We can therefore write the unrenormalized energy as
\begin{eqnarray}
E &=& \frac{1}{\pi} \int_0^\infty k dk \sum_{\ell=1}^\infty (2\ell+1)
\left[\frac{1}{2} \frac{d\delta_{\ell}}{dk} + 
\right. \label{eqn:unrenormalized} \\ && \left.
\int_0^\infty dr \left(\frac{1}{2k} \frac{dV_k(r)}{dk} 
+\epsilon_k(r) - 1\right) |\psi_{k,\ell}(r)|^2\right] \,. \nonumber
\end{eqnarray}
For numerical calculation, this form is greatly preferable to
Eq.~(\ref{eqn:energy}), because it takes advantage of
Eq.~(\ref{eqn:goodwronk}) to yield an expression in which
the $r$ integral has support only on a compact region.
It is the second term in brackets in the $\ell$-summand in
Eq.~(\ref{eqn:unrenormalized}) that leads to a discrepancy between our
results and previous calculations, such as Ref.~\cite{Bordag:2008rc}.

The frequency integral is most conveniently calculated by the variable 
phase method of Ref.~\cite{Graham:2002xq} on the imaginary axis $k=i\kappa$.  
%%
%% sentence added in the revision
%%
It should be stressed again that the continuation to imaginary frequencies 
is not merely a matter of convenience.  Rather it is \textit{mandatory} in view of 
the path integral results of Ref.~\cite{Rahi:2009hm}.
In order to extend the integration to negative momenta, the factor $k$ in 
Eq.~(\ref{eqn:unrenormalized}) must be treated as $\sqrt{k^2}$. This term 
produces the branch cut on the imaginary axis along which we integrate. 
We parameterize (the analytic continuation of) the outgoing wave solution 
via spherical Riccati--Hankel functions as
$$
i\kappa r h_\ell^{(1)}(i\kappa r)\, {\rm e}^{\beta_{\kappa,\ell}(r)}\,,
$$ 
which (for real, positive $k$) gives
$\delta_{\ell}(k)=\lim\limits_{r\to0}
{\sf Im}\left[\beta_{-ik,\ell}(r)\right]$,
where $\beta_{\kappa,\ell}(r)$ obeys
\begin{equation}
-\beta_{\kappa,\ell}''(r) + 
2\kappa \xi_\ell(\kappa r)\beta_{\kappa,\ell}'(r)
 - \beta_{\kappa,\ell}'(r)^2 +  V_{i\kappa}(r) = 0
\label{eqn:beta}
\end{equation}
with the boundary conditions $\lim\limits_{r\to\infty}
\beta_{\kappa,\ell}(r) = 0$ and $\lim\limits_{r\to\infty}
\beta_{\kappa,\ell}'(r) = 0$.  Here $\xi_\ell(\kappa r)$ is given in
terms of spherical Riccati--Hankel functions as
\begin{equation}
\xi_\ell(z) = -\frac{\frac{d}{dz} \left(z h_\ell^{(1)}(i z)\right)}
{z h_\ell^{(1)}(i z)} \,.
\end{equation}
By also parameterizing the regular solution as
\begin{equation}
\psi_{i\kappa,\ell}(r) =
\frac{h_{\kappa,\ell}(r)}{(2\ell+1)(-i\kappa)^\ell
i \kappa r h_\ell^{(1)}(i\kappa r)} \,,
\end{equation}
where $h_{\kappa,\ell}$ obeys
\begin{equation}
- h_{\kappa,\ell}''(r)
- 2 \kappa \frac{d}{dr} \left(\xi_\ell(\kappa r) h_{\kappa,\ell}(r)\right)
+  V_{i\kappa}(r) h_{\kappa,\ell}(r) = 0
\label{eqn:h}
\end{equation}
with $h_{\kappa,\ell}(0)=0$ and $h_{\kappa,\ell}'(0)=1$, we can express the 
norm squared of the wavefunction through the Green's function techniques 
Ref.~\cite{Graham:2002xq} as
\begin{equation}
|\psi_{i\kappa,\ell}(r)|^2 =
\kappa \frac{h_{\kappa,\ell}(r) e^{\beta_{\kappa,\ell}(r)}}{(2\ell+1)
e^{\beta_{\kappa,\ell}(0)}}  \,.
\end{equation}
In the free case, it is given 
in terms of spherical Bessel and Hankel functions by
\begin{equation}
|\psi_{i\kappa,\ell}^{(0)}(r)|^2 =
\kappa^2 r^2 j_\ell(i\kappa r) h_\ell^{(1)}(i\kappa r) \,.
\end{equation}

The energy given by Eq.~(\ref{eqn:unrenormalized}) contains the usual
divergences of quantum field theory and as a result depends on the
ultraviolet cutoff.  We must therefore renormalize the theory in a way
that makes contact with physically measurable quantities.  To do so
requires that we focus attention on the region within the shell
itself, since that is where the local counterterms are nonzero. We
will require two renormalization steps:  one for the leading
quadratic divergence, and a second for the residual logarithmic 
divergence.

We follow the conventional prescription for the leading quadratic 
divergence, the ``no tadpole'' scheme, in which we subtract the leading Born
approximation from the energy
\cite{Graham:2002fw,Graham:2002xq,Graham:2003ib}.  This quantity is
local, that is, proportional to a simple integral over space of the
potential.

The subleading logarithmic divergence emerges from the 
two-point function of the perturbative expansion of QED. Hence 
consistency of the (quantum) Drude model requires a singular local  
counterterm proportional to $\int dr r^2 p(r)^2$. The 
details of this counterterm, in particular its finite terms,
are material properties that would be determined within a microscopic
theory of the dielectric. We avoid those unknowns by only
comparing the vacuum polarization energies of background fields
that have equal square integrals. Our (smooth) profile 
function parameterizes a narrow step-type shape centered at
a particular radius $R$. Within this step the dielectric deviates 
from unity. For such step shapes, holding fixed the integral of any 
power of the profile function yields the same condition on how to 
modify the profile when varying $R$. In particular, constraining 
$\int dr r^2 p(r)$ fixes the number of charge carriers. We use a 
profile function parameterized by a radius $R$ and a steepness $s$,
\begin{equation}
p(r) = \sech \left[s (r-R)\right] \,,
\end{equation}
and then consider the difference between the cases $R = R_1$, 
$s = s_1$ and $R = R_2$, $s = s_2$ such that
$\int_0^\infty r^2 \sech^2 \left[s_1 (r-R_1)\right] dr =
\int_0^\infty r^2 \sech^2 \left[s_2 (r-R_2)\right] dr$.
Our renormalized energy in each channel thus becomes
\begin{align}
E_{\text{\tiny ren}} = 
&\int_0^\infty \frac{d\kappa}{\pi} \mbox{\large $\Delta$} \left\{
\sum_{\ell=1}^\infty (2\ell+1)\left[\frac{1}{2}\left(
\beta_{\kappa,\ell}(0) - \beta_{\kappa,\ell}^{(1)}(0)\right)
\phantom{\int} \right.\right. \nonumber \\ 
& + \left.\left. \kappa^2 \int_0^\infty dr \left(-\frac{1}{2\kappa}
\frac{dV_{i\kappa}(r)}{d\kappa} +\epsilon_{i\kappa}(r) - 1\right)
\right. \right. \label{eqn:renormalized} \\ 
&  \times \left.\left.\left(\frac{h_{\kappa,\ell}(r)
e^{\beta_{\kappa,\ell}(r)}}{(2\ell+1)
e^{\beta_{\kappa,\ell}(0)}} -
\kappa r^2 j_\ell(i\kappa r) h_\ell^{(1)}(i\kappa r)
\right)\right]\right\} \,,
\nonumber
\end{align}
where the first Born approximation $\beta_{\kappa,\ell}^{(1)}(r)$ is
obtained by iteration of the differential equation (\ref{eqn:beta})
and $\Delta\{\ldots\}$ indicates that we compute the difference between 
the cases $R = R_2$, $s = s_2$ and $R = R_1$, $s = s_1$.
%%
%% paragraph added in the revision
%%
The angular momentum sums in expressions like Eq.~(\ref{eqn:renormalized}) 
are finite, and typically converge towards a limiting function of the 
momentum $\kappa$~\cite{Farhi:2001kh}, which can be extracted from 
(local parts of) low-order Feynman diagrams.  The field theory divergences 
all rest within the momentum integral. Our scaling of the profile ensures 
that local parts of those Feynman diagrams, which contain the logarithmic 
divergence, do not vary with $R_2$. Hence they cancel in $\Delta$, yielding 
a finite result.  The sum of this quantity over TE and TM channels then 
gives the total change in energy when expanding or contracting the shell.
We call the term in Eq.~(\ref{eqn:renormalized}) that does not (explicitly) 
involve the radial integral the {\it traditional} contribution, since it 
is the analog of a calculation that is based on the change of the density 
of states measured by the momentum derivative of the phase shift. We call 
the remaining term, which involves the radial integral, the {\it additional} 
contribution.

For imaginary momentum $k=i\kappa$ the dielectric function becomes real, 
as do the potentials, $V_{i\kappa}(r)$, in the wave equations. Rotating 
to the imaginary momentum axis has the further advantage that it allows 
us to change the order of angular momentum sums and linear momentum 
integration~\cite{Schroder:2007xk}. We perform the angular momentum sum 
first, cutting off the numerical computation at $\ell = \ell_{\rm max}$ 
for the sum over angular momentum channels and $\kappa = \kappa_{\rm max}$
for the subsequent (imaginary) momentum integral.  For
$\kappa>\kappa_{\rm max}$ we fit  a power law to the integrand and use
that fit to estimate the contributions from large momenta. This procedure  
requires (i) that $\kappa_{\rm max}$ is large enough to be in the 
asymptotic regime and (ii) that the angular momentum sum up to 
$\ell_{\rm max}$ has converged at $\kappa_{\rm max}$. The convergence 
condition on the angular momentum sum is conditional: the larger $\kappa$, 
the larger we need to take $\ell_{\rm max}$. The condition on $\kappa_{\rm max}$ 
is not very severe, since $\kappa_{\rm max}\sim 2.5s$ turns out to be sufficient.  
However, the angular momentum sum converges slowly and we
need to take  $\ell_{\rm max}$ as big as 2000. Practically, these
conditions can be met for the traditional contribution with
moderate numerical computation, but the additional contribution is 
significantly more costly. For that reason we compute the additional 
contribution for several choices of $\ell_{\rm max}\sim 1000$ and find the 
$\ell_{\rm max}\to\infty$ result by extrapolation. It turns out that the 
contribution from this extrapolation is of similar magnitude in the TE and TM 
channels. But since the former is much smaller than the latter overall, the 
relative effect of the extrapolation is sizable in the TE channel but small 
in the TM channel. 

\begin{figure}[htbp]
\centerline{
\includegraphics[width=0.8\linewidth]{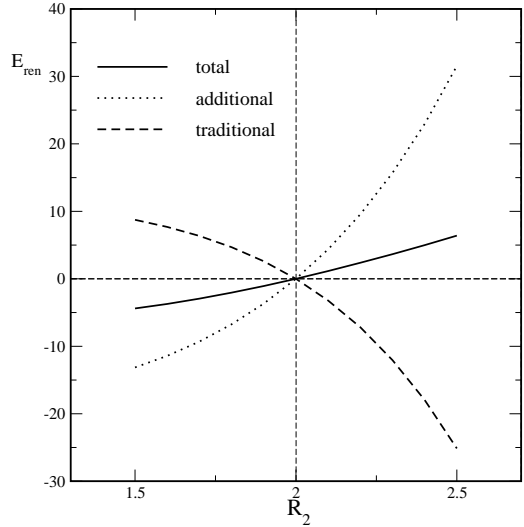}}
\vskip0.3cm
\caption{\label{fig_total} Difference in renormalized energy between
shells of radii $R_2$ and $R_1$, as a function of $R_2$ for $R_1=2$.}
\end{figure}

A further complication arises because we require the Hankel functions
numerically for any order and  any argument.  Standard
algorithms~\cite{nr1992} are not appropriate for very large order and
very small arguments.  To bypass this obstacle, we use those
algorithms only to find $\xi_\ell(z)$ for $z\to\infty$ as a boundary
value, and then solve the non--linear first order differential equation 
\begin{equation}
\frac{d\xi_\ell(z)}{dz}=\xi_\ell^2(z)-1-\frac{\ell(\ell+1)}{z^2}
\label{eq:deqxi}
\end{equation}
to determine the logarithmic derivative of the spherical Riccati--Hankel
function for real $z$. Similarly, we do not use standard algorithms to
find the free Green's function, but rather supplement the set of differential 
equations~(\ref{eqn:h}) by the case $V_{i\kappa}(r)\equiv0$, which yields
$h^{(0)}_{\kappa,\ell}(r)$ with $|\psi^{(0)}_{i\kappa,\ell}|^2=
\frac{\kappa}{2\ell+1}h^{(0)}_{\kappa,\ell}(r)$. The numerical results
presented  here are based on {\sc Fortran} codes. Those programs have
also been tested against {\sc Mathematica} codes for moderate $\ell_{\rm
max}$ and $\kappa_{\rm max}$.
 
In order to make the numerical calculation tractable, we choose moderate 
values of the model and ansatz parameters, rather than attempting to 
closely model a physical metallic sphere. Our results, however, are 
representative of the generic behavior of the stress on a dielectric shell. 
We work in units where $\lambda_p=2$ and choose as our reference contribution 
a shell with $R_1 =2$ and $s_1=4$ in these units.

%%
%% This sentence added in the revision
%% 
Figure~\ref{fig_total} shows the results for the choice $\sigma_p=2$. We have 
also studied larger conductivities ($\sigma_p=4$ and $\sigma_p=8$) and find
the same behavior, indicating that the choice of infra--red regularization 
is not crucial. We find that the traditional contribution, {\it i.e.} the one 
solely based on the phase shift (or equivalently, the logarithm of the Jost 
function~\cite{Bordag:2008rc}), decreases with increasing radii, which would 
indeed lead to a repulsive self--stress  if it were the sole contribution. 
On the other hand, the additional  contribution due to the energy dependence 
of the dielectric, which  is localized at the shell and depends on its material 
properties, is attractive.  In total, the contribution from the additional term 
overcomes the standard repulsion and we find the electromagnetic Casimir stress 
of a spherical dielectric shell to be attractive, in agreement with the generic 
behavior of electromagnetic Casimir forces between rigid bodies, as has been 
established for configurations with mirror symmetry~\cite{Kenneth06} and many 
other geometries (see for example Ref.~\cite{Rahi:2009hm}).

Our result shows how the attractive Casimir stress on a dielectric shell 
arises from a term that is not captured by the idealized boundary 
condition calculation. The additional contribution
clearly originates from the frequency dependence
of the dielectric and therefore is a manifestation of material
properties.  Because the dominant contribution originates in terms in
$V^{\hbox{\tiny TM}}_k(r)$ proportional to space derivatives of
$\epsilon_k(r)$, this result nonetheless persists in any limit in
which the dielectric approaches such a boundary.  We expect this
behavior to apply to other cases for which the ideal boundary suggests
a repulsive stress, such as a rectangular box, again in agreement with
the results from Casimir forces \cite{Hertzberg:2005pr}, although it
is more difficult to formulate the numerical calculation in that case.

\begin{acknowledgments}
N.\ G.\ was supported in part by the National Science Foundation (NSF)
through grant PHY-1213456.  H.\ W.\ was supported by the National
Research Foundation (NRF), Ref. No. IFR1202170025.
\end{acknowledgments}

\end{document}